\RequirePackage{fix-cm}
\documentclass[twocolumn]{svjour3}          
\smartqed  
\usepackage{graphicx}
\usepackage{color}
\usepackage{amssymb}
\usepackage{amsmath}
\usepackage{xspace}

\newcommand{\ie}{\textit{i.e.}\xspace}
\newcommand{\eg}{\textit{e.g.}\xspace}

\renewcommand{\d}{\mathrm{d}}

\newcommand{\poissonratio}{\nu}

\newcommand{\fc}{\ensuremath{\mu}} 
\newcommand{\sfc}{\fc_{\mathrm{s}}} 
\newcommand{\kfc}{\fc_{\mathrm{k}}} 
\newcommand{\slip}{\delta} 
\newcommand{\dc}{d_{\mathrm{c}}} 
\newcommand{\pzone}{w}
\newcommand{\pzoneslope}{\Delta \tau_{\mathrm{pz}}}

\newcommand{\la}{l_{\mathrm{a}}}
\newcommand{\lb}{l_{\mathrm{b}}}
\newcommand{\li}{l_{\mathrm{n}}}

\newcommand{\sigmaN}{\sigma_{\mbox{\tiny{N}}}}
\newcommand{\sigmaS}{\sigma_{\mbox{\tiny{S}}}}
\newcommand{\tauN}{\tau_{\mbox{\tiny{N}}}}
\newcommand{\tauS}{\tau_{\mbox{\tiny{S}}}}
\newcommand{\FN}{F_{\mbox{\tiny{N}}}}
\newcommand{\FS}{F_{\mbox{\tiny{S}}}}

\newcommand{\sif}{K} 
\newcommand{\sifII}{\sif_{\mathrm{II}}} 
\newcommand{\sifIIc}{\sif_{\mathrm{IIc}}}

\begin{document}

\title{Linear elastic fracture mechanics predicts the propagation distance of frictional slip}

\author{David S. Kammer \and Mathilde Radiguet \and Jean-Paul Ampuero \and Jean-Fran\c{c}ois Molinari}

\institute{
  D.S. Kammer  \and M. Radiguet
  \at Computational Solid Mechanics Laboratory, IIC-ENAC, Ecole Polytechnique F{\'e}d{\'e}rale de Lausanne, EPFL, CH-1015 Lausanne, Switzerland
\and
  J.-P. Ampuero
  \at Seismological Laboratory, California Institute of Technology, Pasadena, California, USA
\and 
  J.F. Molinari
  \at Computational Solid Mechanics Laboratory, IIC-ENAC, IMX-STI, Ecole Polytechnique F{\'e}d{\'e}rale de Lausanne, EPFL, CH-1015 Lausanne, Switzerland, \\\email{jean-francois.molinari@epfl.ch}
}

\date{\today}

\maketitle

\begin{abstract}
When a frictional interface is subject to a localized shear load, it is often (experimentally) observed that local slip events initiate at the stress concentration and propagate over parts of the interface by arresting naturally before reaching the edge.
We develop a theoretical model based on linear elastic fracture mechanics to describe the propagation of such precursory slip. The model's prediction of precursor lengths as a function of external load is in good quantitative agreement with laboratory experiments as well as with dynamic simulations, and provides thereby evidence to recognize frictional slip as a fracture phenomenon. We show that predicted precursor lengths depend, within given uncertainty ranges, mainly on the kinetic friction coefficient, and only weakly on other interface and material parameters. By simplifying the fracture mechanics model we also reveal sources for the observed non-linearity in the growth of precursor lengths as a function of the applied force. The discrete nature of precursors as well as the shear tractions caused by frustrated Poisson's expansion are found to be the dominant factors. Finally, we apply our model to a different, symmetric set-up and provide a prediction of the propagation distance of frictional slip for future experiments.
\end{abstract}

\section{Introduction}
\label{sec:introduction}

\begin{figure}
  \includegraphics[width=0.99\columnwidth]{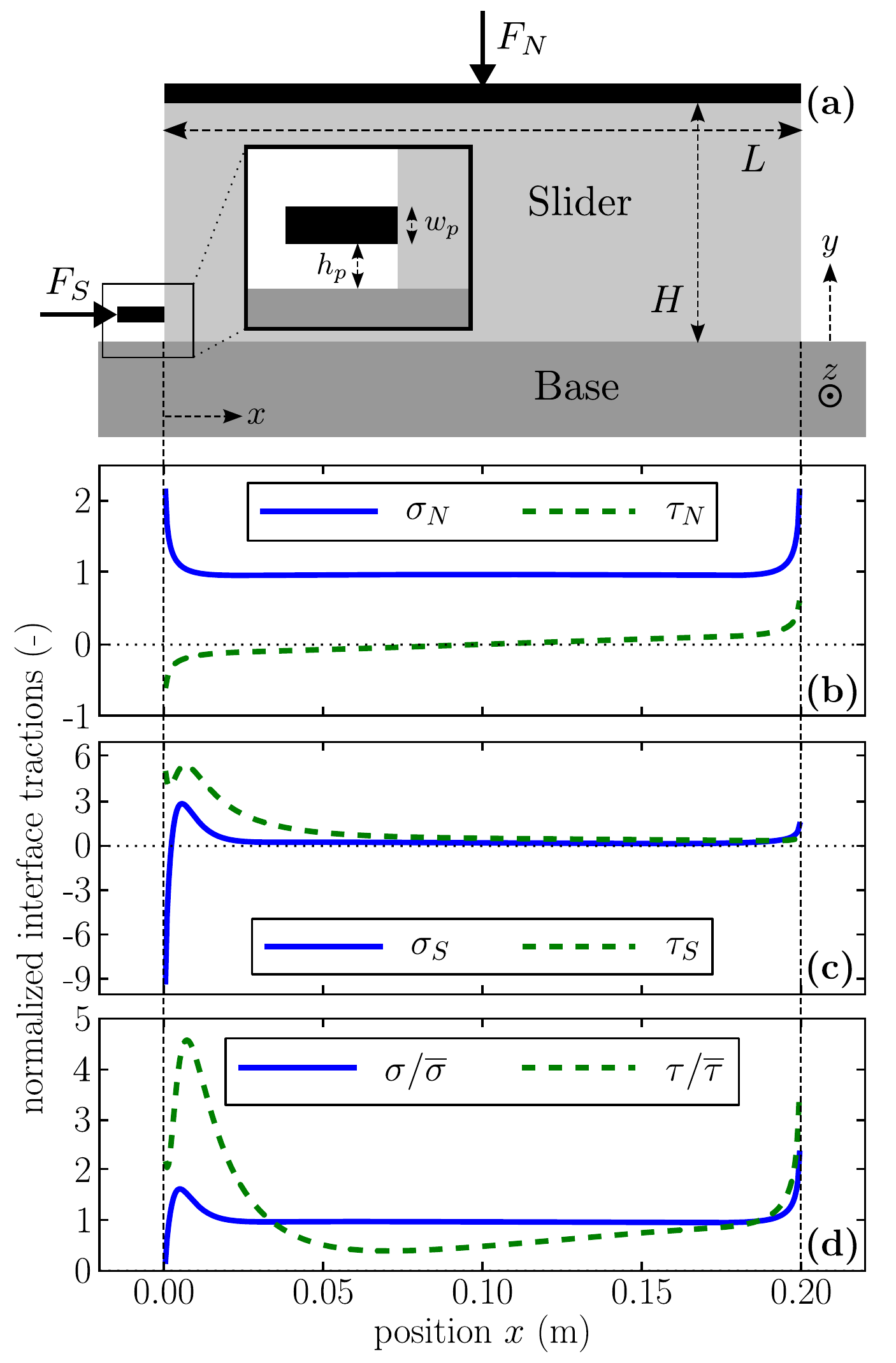}
  \caption{(a) Set-up of the side-driven system. A thin slider is pressed by a constant normal load $\FN$ onto a thicker base block. A pusher applies a slowly increasing shear load $\FS$ to the slider. (inset) Zoom on the pusher. (b) Static stress state at the interface for an applied normal load. Note the convention of positive $\sigma$ for compression. (c) Static stress state at the interface for an applied unit shear load. (d) Interface stresses for an unruptured linear combination of both loadings with $\FN = 5 \FS$ and normalized by the average interface stress.}
  \label{fig:initial_stresses}
\end{figure}

\begin{figure*}
  \includegraphics[width=0.99\textwidth]{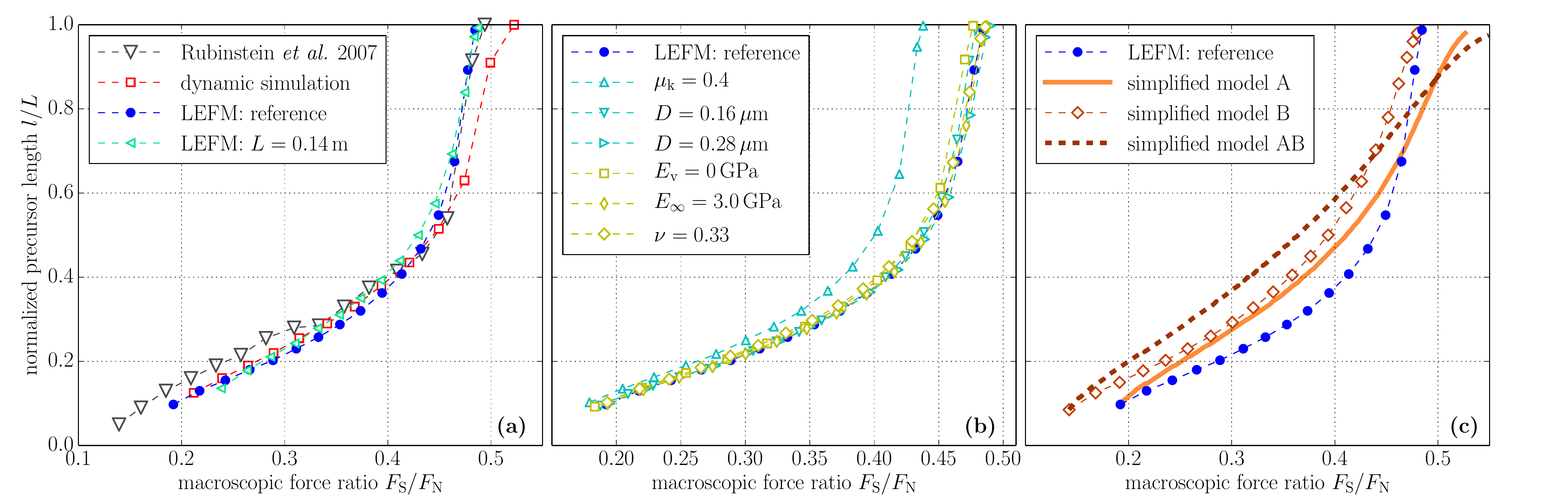}
  \caption{Evolution of normalized precursor length with increasing macroscopic force ratio $\FS/\FN$. (a) Comparison of LEFM theory with experimental data from~\cite{rubinstein:2007}, and dynamic finite-element simulations from~\cite{radiguet:2014}. 
(b) Influence of different interface and material parameters. Parameters that are changed with respect to the reference case are given in the legend. The variation of the equivalent slip distance $D$ for $\kfc = 0.45$ corresponds to the uncertainty range of the frictional fracture energy deduced from experiments~\cite{svetlizky:2014}. 
(c) Comparison of full LEFM theory with simplified models. Model A: traction changes due to interface ruptures are neglected [$\Delta \tau_i(x) = 0$]. Model B: Interface shear tractions due to frustrated Poisson's expansion are neglected [$\tauN (x) = 0$]. Model AB: combination of model A and B [$\Delta \tau_i(x) = 0$ and $\tauN (x) = 0$].
}
  \label{fig:results}
\end{figure*}

Recent laboratory experiments have shown that nominally flat interfaces between solids under a localized quasi-static shear load may present local slip precursors well before global sliding~\cite{rubinstein:2007,maegawa:2010}. These findings on the transition from sticking to sliding have attracted wide attention~\cite{scheibert:2010,tromborg:2011,bouchbinder:2011,amundsen:2012,kammer:2012,otsuki:2013}. They have important implications in engineering as well as earthquake science, where spatially concentrated loads appear at the base of most faults and ruptures propagate over parts of the interfaces~\cite{lapusta:2003,wu:2014}. In the experiments, two PMMA (acrylic glass) blocks are brought into contact under a constant normal load $\FN$. A shear load $\FS$ is applied to the top block (slider) via a pusher located close to the interface [Fig.~\ref{fig:initial_stresses}(a)]. In this side-driven set-up, local slip fronts nucleate episodically at the trailing edge and propagate over parts of the interface. 
Their propagation distance increases proportionally to the applied load until approximately the middle of the interface. From this point on, the growth of the precursors is considerably faster [\eg, see experimental data from~\cite{rubinstein:2007} shown in Fig.~\ref{fig:results}(a)]. Once a slip event propagates over the entire interface, global sliding occurs.

The remarkable increase of precursor lengths and its non-linear relation to the applied shear force $\FS$ was shown to be highly reproducible and, if normalized by sample length and normal force, unique and independent of the slider geometry (length or height), of the normal load, and of the pusher position~\cite{rubinstein:2007,tromborg:2011}. 
In essence, episodic nucleation and spontaneous arrest of precursor fronts arise from the spatial concentration of interface stresses induced by the applied load~\cite{rubinstein:2008}.
Several numerical models, one-dimensional spring-mass chains with arbitrary normal loads~\cite{maegawa:2010,amundsen:2012} as well as two-dimensional spring-mass models~\cite{tromborg:2011}, were proposed to simulate the mechanics of precursors and to analyze the relation between the normalized precursor length and the measured macroscopic force ratio $\FS / \FN$.
They confirmed experimental observations showing the absence of influence of the slider geometry and produced non-linear evolution of the precursor length. However, none of these numerical simulations provided quantitative comparison to experimental data. The main reasons being inconsistent interface stresses due to the discrete nature~\cite{maegawa:2010,tromborg:2011,amundsen:2012} or the one-dimensional geometry~\cite{maegawa:2010,amundsen:2012} of these models.

In addition to numerical models, only few theoretical approaches have been proposed so far. A quasi-static one-dimensional model~\cite{scheibert:2010} applied a simplified stress criterion, inspired by Griffith's energetic criterion, to study the kinematics of the transition from static to stick-slip friction, and showed that it is dominated by the system instead of the small scale parameters. Another analytical model~\cite{amundsen:2012} applied a reverse approach. Given a precursor length, they describe the interface stresses after precursor arrest and compute the associated macroscopic shear force by integration of the interface shear tractions.
Even though these theoretical models offer interesting intuition about the propagation length of precursors, they do not give new insights about the mechanics of friction nor do they provide quantitative comparison with experiments. 
There are several causes to this discrepancy. As for the numerical simulations, the one-dimensional geometry and the arbitrary initial stress states can result in inconsistent predictions. Moreover, the simplistic propagation criteria, and the use of post-precursor instead of pre-precursor stress states are additional limitations of existing models.

In this Letter, we present a theoretical model based on Linear Elastic Fracture Mechanics (LEFM) that predicts the kinematics of slip precursors at frictional interfaces. 
Such LEFM approaches have long been used in earthquake modeling~\cite{freund:1979,ampuero:2006,kato:2012} and recently, 
experimental evidence was provided by measurements of LEFM strain fields around  Sub-Rayleigh slip fronts~\cite{svetlizky:2014}. 
We here develop this concept into a quantitative model that incorporates the continuum nature of fracture mechanics theory as well as interface stress states resulting from the exact system geometry.
The aim of this model is to be as simple as possible while comprising the essential features of fracture mechanics theory (more precisely LEFM) and addressing the shortcomings of previous theoretical models.
With this approach, we study the link between meso-scale properties and the macro-scale response of a solid-body system containing a frictional interface. Friction mechanisms acting at even smaller scales (\ie, atomic scale) are incorporated in a local (meso-scale) friction law.

Specifically, we use real interfacial stress states from a two-dimensional geometry to provide a prediction of the precursor length based on the shear load measured before the slip event and compare our results quantitatively with experimental data.
In addition, we analyze the influence of several material and interface parameters, which has not been done before, and point out various sources of the precursor length non-linearity.
This analysis is further extended by considering simplifications of our model which provide a fundamental understanding of the origin of the non-linearity of the precursor length evolution.

\section{Model}
\label{sec:model}

As in~\cite{rubinstein:2007}, the studied system is modeled by a rectangular thin plate of length $L = 200\,$mm, height $H = 75\,$mm and thickness $b = 6\,$mm, in contact with a much thicker deformable base block of dimensions $300 \times 30 \times 27\,$mm. A pusher of width $w_p = 5\,$mm is applied at height $h_p = 6\,$mm from the interface. The material properties are assumed to be viscoelastic,
with Poisson's ratio $\poissonratio$, viscous $E_{\mathrm{v}}$ and static $E_{\infty}$ Young's moduli. The resulting instantaneous Young's modulus is given by $E_0 = E_{\mathrm{v}} + E_{\infty}$.
A linear slip-weakening friction law~\cite{palmer:1973,andrews:1976} is applied at the interface, describing the frictional strength as
\begin{equation}
  \tau^s (\slip, x) = \max \left( \kfc , ~  \sfc + \slip / \dc \left( \kfc - \sfc \right)\right) \sigma (x)~,
  \label{eq:frictionlaw}
\end{equation}
where $\sfc$ and $\kfc$ are the static and kinetic friction coefficients, $\dc$ is the characteristic weakening length, $\slip$ is the local interface slip, $\sigma (x)$ is the contact pressure, and $x$ is the coordinate along the interface. 
More advanced friction laws, such as velocity-weakening-strengthening friction, have been used in the past to model precursor mechanics at PMMA interfaces~\cite{braun:2009,bouchbinder:2011,kaneko:2011,bar-sinai:2012,bar-sinai:2014}. Even though these models describe well the propagation of frictional slow fronts, they are not indispensable to model the propagation distance of precursors, as shown with dynamic finite-element simulations using slip-weakening friction~\cite{radiguet:2013}. Here, the emphasis is on simplicity and the slip-weakening friction law enables simple determination of the interface's fracture toughness, which is essential to LEFM theory.

The unruptured tractions at the interface are computed by static finite-element simulations. An applied unit normal load leads to a normal $\sigmaN (x)$ traction satisfying the following condition $b \int_0^L \sigmaN (x) \d x = 1\,$N. The resulting contact pressure [Fig.~\ref{fig:initial_stresses}(b)] is approximately uniform in the central $80\%$ of the interface and presents singularities at the edges due to the perfect rectangular shape of the specimen. Poisson's lateral expansion is frustrated at the interface by the frictional strength leading to a shear traction $\tauN (x)$, which is approximately linear and symmetric with respect to the center point of the interface. 
Similarly, $\sigmaS (x)$ and $\tauS (x)$ result from an applied unit shear load and satisfy therefore  $b \int_0^L \tauS (x) \d x = 1\,$N. As shown in Fig.~\ref{fig:initial_stresses}(c), $\tauS (x)$ presents a maximum close to the trailing edge, which will eventually lead to the initiation of precursors.

It is important to note that although the unruptured interface tractions are computed by static finite-element simulations, the following model is theoretical and independent of numerical simulations. Any interface stress state, also experimental data, could be used as starting point for our model.

Once the unruptured interface tractions, caused by external loadings, are known, 
the effective interface tractions are then modeled by linear superposition of these tractions and the stress drops due to previous interface ruptures. 
The normal $\sigma_r (x)$ and shear $\tau_r (x)$ tractions after $r-1$ precursors, for any $\FN$ and $\FS$, and after viscous relaxation are given by
\begin{align}
  \label{eq:sigma_r}
  \sigma_r (x) &= \tilde \FN \sigmaN (x) + \tilde \FS \sigmaS (x) \\
  \tau_r (x) &= \tilde \FN \tauN (x) + \tilde \FS \tauS (x) + \frac{E_{\infty}}{E_{0}} \sum_{i = 1}^{r-1} \Delta \tau_i (x) ~,
  \label{eq:tau_r}
\end{align}
with $\tilde \FN$ and $\tilde \FS$ ensuring that the macroscopic normal and shear loads are always equal to $\FN$ and $\FS$, \eg, $b \int_0^L \sigma_r (x) \d x = \FN$ and $b \int_0^L \tau_r (x) \d x = \FS$.
The change in the shear tractions caused by interface rupture $i$ is introduced as $\Delta \tau_i (x)$, while contact pressure changes are neglected.
Taking into account the effect of the bulk material's visco-elasticity, the stress drops have to be multiplied by $E_{\infty}/E_0$ if full relaxation occurs between two slip events~\cite{radiguet:2013,radiguet:2014}.
Furthermore, a non-adhesion condition defines that where $\sigma_r(x) < 0$ we impose: $\sigma_r(x) = 0$ and $\tau_r(x) = 0$. 
An example of an effective stress state without a stress drop is shown in Fig.~\ref{fig:initial_stresses}(d). The contact pressure is rather uniform, while the shear traction presents an important peak close to the trailing edge, which is at the origin of slip nucleations.

\begin{figure}
  \includegraphics[width=0.95\columnwidth]{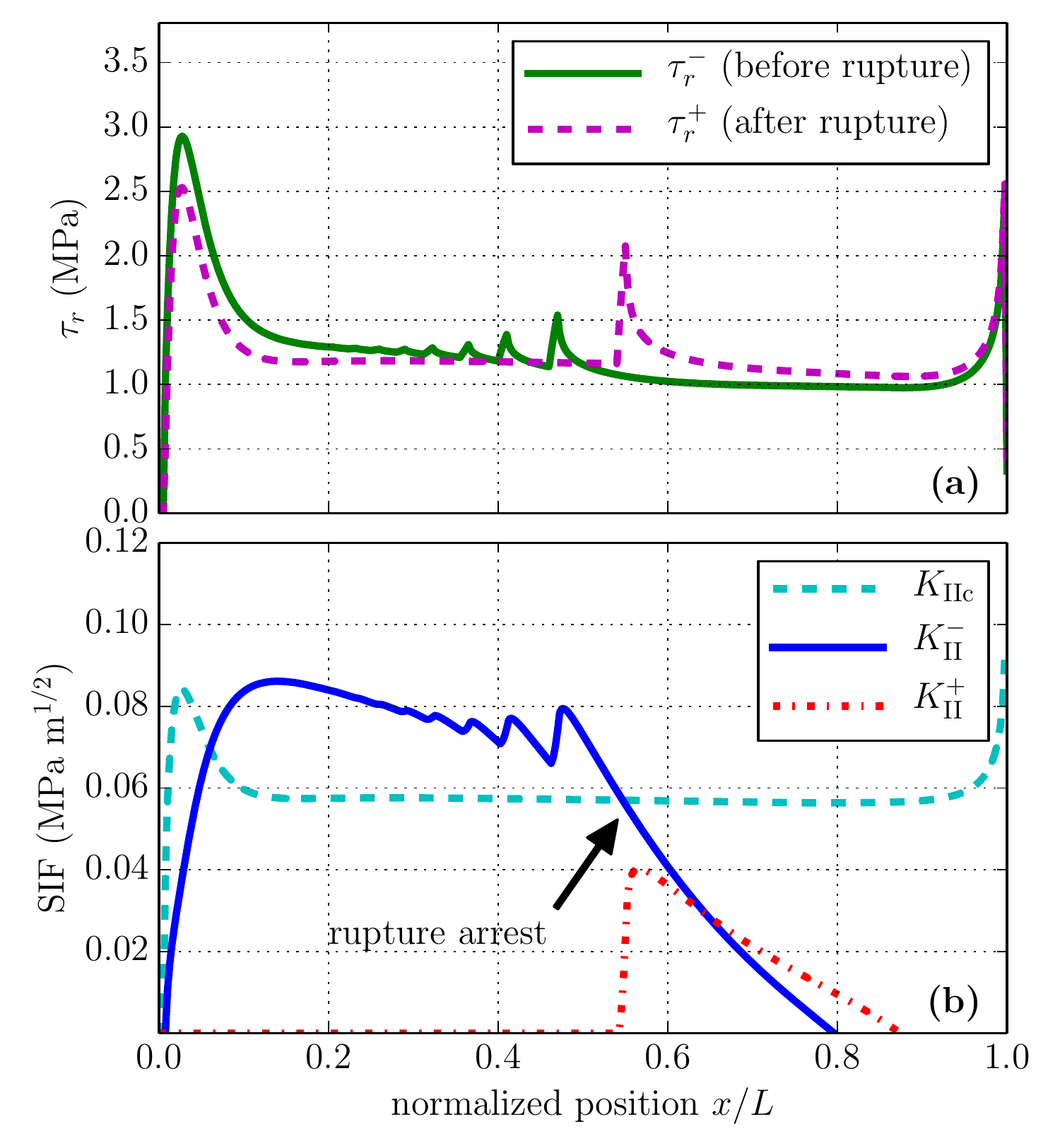}
  \caption{(a) Interfacial shear tractions before and after an interface rupture, given by Eq.~\ref{eq:tau_r} for $r=13$. (b) Fracture toughness $\sifIIc$ and stress intensity factor $\sifII^-$ (before) and $\sifII^+$ (after) the rupture shown in (a).}
  \label{fig:method}
\end{figure}

Considering local slip events as interface ruptures, we model their propagation using LEFM~\cite{book:freund:1990}, which implies that every rupture modifies the stress state of the interface behind as well as ahead of its tip. An example of how shear tractions change during a slip event is shown in Fig.~\ref{fig:method}(a) for a rupture with arrest position $x/L = 0.55$. The shear tractions before and after an interface rupture (in time) are denoted with a superscript $-$ and $+$, respectively. 
The arrest of the precursor creates a peak at $x/L = 0.55$, and a square root decrease in shear tractions for $x/L > 0.55$ [see $\tau_r^+$]. The peaks in $\tau_r^-$ at $x/L = 0.3-0.5$ are the remains of stress concentrations of previous precursors. After the current rupture, they are erased due to the linear slip-weakening friction law [see $\tau_r^+$] and will partially reappear over time. This effect was shown to be the result of the bulk's visco-elasticity~\cite{radiguet:2013,radiguet:2014}. 

The tractions before and after the rupture $r$ are linked by the stress change $\Delta \tau_r (x)$ through $\tau_r^+ (x) = \tau_r^- (x) + \Delta \tau_r (x)$.
The stress $\tau_r^-$ is equal to $\tau_r(x)$ (Eq.~\ref{eq:tau_r}) for $\FN$ and $\FS$ at the time of the rupture.
The stress $\tau_r^+ (x)$ is the result of the rupture and can be separated into three different areas as described below.

At the rupture tip, there is a process (weakening) zone, where $\slip < \dc$ and in which the shear traction drops from the static frictional strength $\tau^s(0,x)$ to the kinetic strength $\tau^s(\dc, x)$. A linear slip-weakening friction law results within the process zone in a non-linear shear traction distribution, which, for reasons of simplicity, is here approximated by a linear function. The size of a static linear process zone is given by $\pzone = 9 \pi \sifII^2(l) / [32 \, \sigma_r^2(l) \, (\sfc - \kfc)^2]$ \cite{palmer:1973}, where $l$ is the arrest position of the precursor and $\sifII$ the mode II stress intensity factor. The leading end of the process zone is at $x = \la$ and the trailing end at $x = \la - \pzone = \lb$. The position $\la$ of the leading end is determined by the stress concentration as defined below, which always results in a process zone that satisfies $\lb < l < \la$.

Behind the process zone, the stress state is imposed by the friction law (Eq.~\ref{eq:frictionlaw}). Because $\slip > \dc$ everywhere, we can write $\tau_r^+ (x) = \tau^s (\dc, x) = \kfc \sigma_r(x)$ for $x<\lb$.

Ahead of the slip event appears a stress concentration caused by the stress drop occurring behind the rupture tip. The stress change ahead is given in first order approximation as $\Delta \tau_r (x) \approx \sifII(l) / \sqrt{2 \pi (x - l)}$. Because frictional rupture does not allow for stress singularities, the frictional strength limits the maximal shear traction, similar to the assumption of a small plastic zone size in fracture mechanics. Therefore the position of the leading end of the process zone is determined such that $\sfc \sigma_r(\la) = \tau_r^-(\la) + \sifII(l) / \sqrt{2 \pi (\la - l)}$. 

This is only a simplified approximation to the correct description of the stress state around a cohesive crack. In fact, the details have no significant effect on the precursor load-length relation studied here, and even neglecting entirely the process zone results in virtually the same observations with isolated shorter slip events that do not affect the load-length relation of the expanding precursors.

The stress change caused by an interface rupture can therefore be summarized as
\begin{equation}
  \Delta \tau_r (x) = \left\{
    \begin{array}{l l}
      \frac{\sifII(l)}{\sqrt{2 \pi (x - l)}} & \textrm{for} \quad x \geq \la \\
      \Delta \tau_r(\lb) + \frac{x - \lb}{\pzone} \, \pzoneslope & \textrm{for} \quad \lb < x < \la \\
      \kfc \sigma_r(x) - \tau_r^-(x)               & \textrm{for} \quad x \leq \lb ~,\\
    \end{array} \right.
\label{eq:stress_drop}
\end{equation}
with $\pzoneslope = \Delta \tau_r(\la) - \Delta\tau_r(\lb)$. The process zone is characterized by $l$ the arrest position of the rupture, $\la$ and $\lb$ the leading and trailing end, respectively, and $\pzone = \la - \lb$ the process zone size.

The mode II stress intensity factor for a non-uniform shear stress drop $\Delta \tau_r$ along an edge crack of length $a$ in a semi-infinite solid can be deduced from Equation 8.3 in~\cite{book:tada:2000} by integration:
\begin{equation}
\label{eq:stress_intensity_factor}
\sifII (a) = \frac{2}{\sqrt{\pi a}}\int_0^a \frac{\Delta \tau_r (s) F(s/a)}{\sqrt{1-(s/a)^2}} \d s
\end{equation}
with $F(s/a) = 1 + 0.3 (1 - (s/a)^{5/4})$ and $\Delta \tau_r (s) = \kfc \sigma_r(s) - \tau_r^- (s)$ because the integration is along the crack interface and the process zone is neglected. A different possible choice of stress intensity factor is a semi-infinite crack approaching the edge of a semi-infinite solid (Equation 9.5 in \cite{book:tada:2000}). On the studied system, this stress intensity factor leads to an almost identical precursor load-length relation as in the model with Eq.~\ref{eq:stress_intensity_factor}. Only a slightly steeper curve at $l/L > 0.5$ is observed (not shown here).  As $\Delta \tau_r$ is multiplied by the non-linear factor $F(s/a)/\sqrt{1 - (s/a)^{2}}$ over the crack face, the stress intensity factor is one possible source of non-linearity in precursor mechanics.

Given that the slider is a thin plate, the fracture toughness is computed in the plane-stress approximation with the frictional fracture energy $G$ by:
\begin{equation}
  \sifIIc (x) = \sqrt{E_0 ~ G(x)} = \sqrt{E_0  \frac{(\sfc - \kfc) \dc}{2} \sigma_r(x)} ~.
\label{eq:fracture_toughness}
\end{equation}
The fracture toughness is computed using $E_0$ because the characteristic frictional weakening time is significantly smaller than the relaxation time of the viscoelastic material \cite{rice:1979}.

Neglecting any dynamic effect, the precursor length $l$ for a given stress state of the interface is determined by the position at which the stress intensity factor becomes smaller than the fracture toughness:
\begin{equation}
  \sifII^- (l) = \sifIIc (l) \quad \textrm{and} \quad \frac{\d \sifII^- (l)}{\d x} < \frac{\d \sifIIc (l)}{\d x} ~.
\label{eq:precursor_length}
\end{equation}
An example is shown in Fig.~\ref{fig:method}(b).
The stress intensity factor right after an event $\sifII^+$ is significantly lower than $\sifIIc$, hence a finite load increment is required to nucleate the next precursor event.

Up to this point, we presented how the precursor length can be predicted for any given interface stress state. In order to complete the proposed model, we need to determine the shear force at which a slip event is expected. As the initiation of the rupture occurs at the trailing edge of the system and a rupture only propagates where the stress intensity factor is larger than the fracture toughness, we introduce a length scale $\li$ which represents the size of the nucleation zone and define that the next precursor occurs when the following condition is satisfied:
\begin{equation}
  \sifII^- (\li) = \sifIIc (\li) \quad \textrm{and} \quad \frac{\d \sifII^- (\li)}{\d x} > \frac{\d \sifIIc (\li)}{\d x} ~.
\label{eq:initiation_condition}
\end{equation}
The slip nucleation zone size $\li$ acts like a seed crack to the propagation of an interface rupture and can be thought of as the stable slip zone that occurs before dynamic ruptures \cite{uenishi:2003,garagash:2012}. Its size may vary from one to another slip event, but is chosen to be constant in our model. However, testing different values for $\li$ has shown that below a critical length, it has only a negligible influence on the precursor load-length relation. Decreasing $\li$ only leads to slightly less precursors. In this work, we chose $\li = 0.012\,$m, which is below the critical length and results in approximately the same precursor occurrence frequency as in the experiment of \cite{rubinstein:2007}. 

\begin{figure}
  \includegraphics[width=0.95\columnwidth]{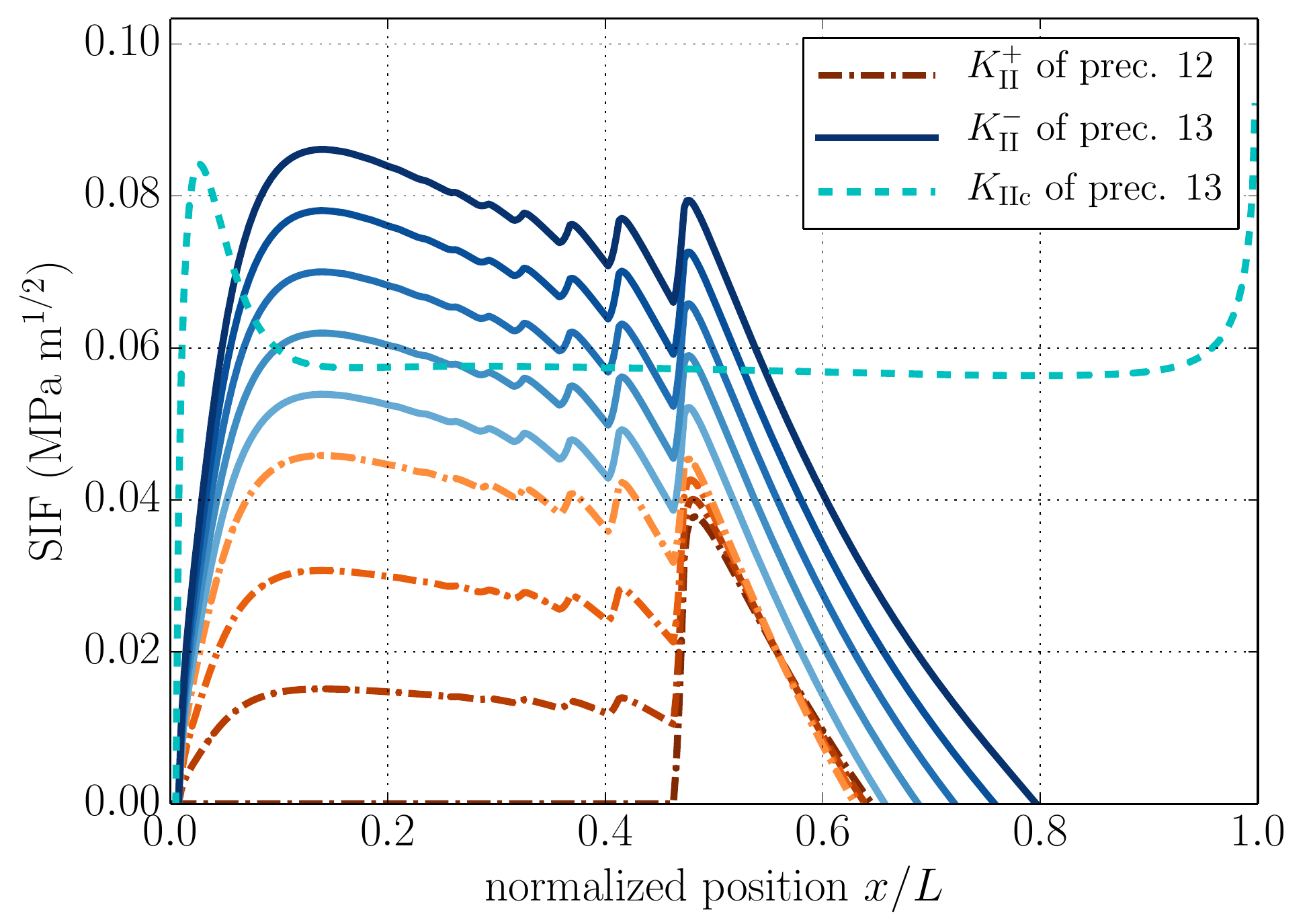}
  \caption{The evolution of the stress intensity factor is shown during the period between two precursors. Directly after precursor $12$, the stress intensity factor $\sifII^+$ is zero along the interface and non-zero ahead of the arrest position, as shown by the dash-dotted dark orange curve. Viscous relaxation of the bulk material, illustrated by the dash-dotted orange curves going from dark to bright, leads to a partial recovery of the pre-rupture stress intensity factor. Further, the increasing external shear load lifts continuously the stress intensity factor, as shown by the solid blue curves going from bright to dark. When the area of $\sifII^- > \sifIIc$ reaches the slip nucleation zone characterised by $x \leq \li = 0.06L$, the next precursor propagates. For simplicity, the effects of the viscous relaxation and the external loading are here illustrated sequentially. In reality, they occur simultaneously. However, if complete relaxation occurs between two precursors, the sequential and simultaneous approaches are equivalent.}
  \label{fig:cycle}
\end{figure}

Before comparing our model with experimental data and studying the influence of various parameters, we here summarize the events occurring during a cycle of an interface rupture in order to provide the reader with a basic intuition of the observed phenomenon. 
Considering an interface stress state at which a slip event occurs [\eg Fig.~\ref{fig:method}(a)], a rupture propagates from the trailing (left) edge until a point where the stress intensity factor becomes smaller than the fracture toughness [Eq.~\ref{eq:precursor_length} and Fig.~\ref{fig:method}(b)]. Behind the rupture occurs a stress drop and ahead of the tip a stress concentration as described by Eq.~\ref{eq:stress_drop}. The stress concentrations of previous ruptures are erased because behind the process zone the friction law imposes shear tractions that depend only on the kinetic friction coefficient and the contact pressure. The viscous memory effect of the bulk material restores these concentrations partially over time \cite{radiguet:2013}. 
Directly after the rupture and before viscous relaxation, the stress intensity factor is zero along the interface up to the arrest position [see $\sifII^+$ in Fig.~\ref{fig:method}(b) and Fig.~\ref{fig:cycle}]. Thus, additional external shear loading is needed to reach a new interface stress state that allows for the propagation of a slip event. While the external loading increases, the stress intensity factor exceeds the fracture toughness first, for this particular set-up, at approximately $x/L = 0.15$ and short after at a position close to the last arrest position (see Fig.~\ref{fig:cycle}). Nevertheless, no rupture initiates because the shear traction is still below the static strength, $\tau_r(x) < \tau^s(0,x)$ (at the last arrest position due to viscous relaxation), 
and the stress intensity factor should be higher than the fracture toughness starting from the edge (and not solely in the middle of interface).
For even higher external shear loads, the area with $\sifII^- > \sifIIc$ expends and once it reaches the seed crack at the edge, and satisfies Eq.~\ref{eq:initiation_condition}, a new slip event occurs and the cycle starts over again.

\section{Results and Discussion}
\label{sec:results_and_discussion}

\subsection{Comparison to Experimental Data}
\label{sec:comparison}

In Fig.~\ref{fig:results}(a), we compare the LEFM prediction (blue dots) with experimental data from~\cite{rubinstein:2007} (gray triangles). 
Material parameters correspond to PMMA~\cite{ciccotti:2004} and interface parameters are deduced from experimental measurements~\cite{rubinstein:2007,ben-david:2010a,svetlizky:2014}: 
$E_{\infty} = 2.6\,$GPa, $E_{\mathrm{v}} = 3.0\,$GPa, $\poissonratio = 0.37$, $\sfc = 0.9$, $\kfc = 0.45$, $\dc = 1\,\mu$m, and $\FN = 3300\,$N. 
The LEFM prediction is in good quantitative agreement with experimental data and retrieves well the non-linearity of the length vs. load curve. 
It is also in good quantitative agreement with results from dynamic finite-element simulations presented in~\cite{radiguet:2014} (Fig.~\ref{fig:results}(a), red squares). These simulations are in plane-stress (slider) and plane-strain (base) approximation with the same geometry, material and interface parameters as for the LEFM model (for more details see~\cite{radiguet:2014}). 
Further, we also confirm the observation of~\cite{tromborg:2011} that the slider geometry does not influence the normalized precursor length behavior by changing the slider length to $L=0.14\,$m [see cyan triangles in Fig.~\ref{fig:results}(a)].

We present in Fig.~\ref{fig:results}(b) the influence of various material and interface properties with variations of the order of their uncertainties.
The value of $\kfc$ is estimated by the macroscopic force ratio $\FS / \FN$ measured directly after a slip event, which is often $\kfc \approx 0.4 - 0.45$~\cite{rubinstein:2007}. 
Introducing an equivalent slip distance $D = \left(\sfc - \kfc\right)\dc/2$ enables us to write Eq.~\ref{eq:fracture_toughness} as $ \sifIIc (x) = \sqrt{E_0 ~ D ~ \sigma(x)}$. According to the experiment-based estimation of the frictional fracture energy of PMMA interfaces reported in~\cite{svetlizky:2014}, the uncertainty of the equivalent slip distance can be determined to be within the range of $D = 0.22 \pm 0.06\,\mu$m for $\kfc = 0.45$. 

With the exception of $\kfc$, the variations of all material and interface parameters within their uncertainties have negligible effects on the $l/L \, - \, \FS/\FN$ relation [Fig.~\ref{fig:results}(b)]. Even neglecting entirely the viscoelasticity of the bulk material ($E_{\mathrm{v}} = 0\,$GPa) does not affect the precursor behavior. This weak influence originates from the square root contribution of $E_0$ and $D$ to $\sifIIc$.
Only a change of $\kfc$ within its uncertainty range results in an important shift of the  $l/L \, - \, \FS/\FN$ curve due to its additional contribution to $\sifII$ [Eq.~\ref{eq:stress_drop} and Eq.~\ref{eq:stress_intensity_factor}].

\subsection{Test of Model Assumptions}
\label{sec:test_of_model_assumptions}

We have shown that the LEFM model is able to produce an accurate prediction of the precursor load-length curve, and reproduce the transition from the initial linear length increase to faster increase at a finite value of load. In the following, we aim at giving a more fundamental understanding of the origin of the load-length curve, by identifying several sources of the non-linearity in this scaling. This is done by removing different components from the LEFM model.

We present in Fig.~\ref{fig:results}(c) a simplification of the theoretical model (denoted model A), which is based on the same LEFM approach, but where any change of the interface tractions due to slip is neglected ($\Delta \tau_i = 0 \quad \forall i$). Under these conditions, the discrete nature of precursors is lost. The length associated to a given macroscopic force ratio is independent of the slip history of the interface, and corresponds to the length that the first precursor would reach if it initiated at that specific loading. The loss in discreteness results in a more (but still not) linear $l/L \, - \, \FS/\FN$ relation. In this simplified model, for a given $\FS$, the stress drop close to the trailing edge is larger than in the reference case, resulting in a higher value of $\sifII$ and in longer precursors. 
But for $l/L > 0.7$, this effect is compensated in the full theory by the stress redistribution close to the arrest position of the previous precursor.

As noted before, the shear tractions at the interface result not only from the macroscopic shear load but also, due to frustrated Poisson's expansion, from the normal load. The influence of the latter is analyzed in model B. The shear contribution of the macroscopic normal load is removed by setting artificially $\tauN(x) = 0 \quad \forall x$.
All remaining interface tractions are kept the same as for the reference case. 
The resulting propagation distances reported with respect to the macroscopic force ratio are shown in Fig.~\ref{fig:results}(c). For any given shear force $\FS$, the precursor length is longer for the simplified model B than for the reference model. For $l/L < 0.5$, this is the logic consequence of neglecting $\tauN(x)$ which acts against the driving traction $\tauS (x)$. Beyond the central point of the interface, the precursor lengths increase faster but still less than in the reference system, where $\tauN(x)$ contributes to the propagation of precursors. From a global perspective, the precursor load-length curve is still non-linear (but less than the reference model). This indicates that the interfacial shear traction resulting from frustrated Poisson's expansion is one but not the only source of non-linearity in the system. 

We also present the results of the simplified model AB, which is the combination of model A and B, where stress drops due to interface ruptures as well as shear traction caused by frustrated Poisson's expansion are neglected. As for model A, the discrete nature of precursors is lost in model AB. The precursor load-length relation, which is shown in Fig.~\ref{fig:results}(c), is almost perfectly linear indicating that most sources of non-linearity (at least for the studied system and parameter range) are eliminated from this simplified model. The non-linear form of the stress intensity factor, which is still part of model AB, does not seem to affect the precursor propagation distance much within the length of the interface.

\subsection{Insights from a Minimalistic Model}
\label{sec:simplistic_model}

In the previous section, we simplified the LEFM model by removing different components in order to analyze their contributions to the non-linearity of the precursor load-length relation. In this section, we apply fracture mechanics in an even simpler model. 

Let us assume, in order to simplify the computation of the stress intensity factor, that the edge crack considered so far is half of a central shear crack at a weak interface of length $2L$. The interface is subjected to a linear shear load and to a point shear load at the center of the crack. The linear shear load corresponds in the full model to the effect of the frustrated Poisson's expansion. The point load represents the localized shear load caused by the pusher.
The stress drop along this interface is therefore given by
\begin{equation}
  \Delta \tau \left( x \right) = \frac{2 \tauN^{\mathrm{max}}}{L} \left( |x| - \frac{L}{2} \right) + \FS \, \delta(x) - \tau_{\mathrm{d}} ~,
  \label{eq:delta_tau_dirac}
\end{equation}
where $\tauN^{\mathrm{max}}$ is the maximal shear traction (at $x=L$) due to frustrated Poisson's expansion, $\FS$ is the amplitude of the point load, $\delta(x)$ is the Dirac delta function, and $\tau_{\mathrm{d}}$ is the dynamic shear stress left behind the crack.

The stress intensity factor of a central crack of length $2a$ is found by integration of equation 5.11 of \cite{book:tada:2000}:
\begin{align}
  \sifII \left( a \right) &= 2\sqrt{\frac{a}{\pi}} \int_0^a \frac{\Delta \tau(s)}{\sqrt{a^2 - s^2}} \d s \\
  &= \sqrt{\pi a} \left[ \tauN^{\mathrm{max}} \left(\frac{4}{\pi}\frac{a}{L} - 1\right) + \frac{2}{\pi} \frac{\FS}{a} - \tau_{\mathrm{d}} \right] ~.
  \label{eq:central_crack_sif}
\end{align}
Using the same propagation criterion to predict precursor length $l$ as for the full theory, we can write $\sifII(l) = \sifIIc(l)$, which leads to
\begin{equation}
  \sqrt{\frac{l_{\mathrm{c}}}{L}} = \sqrt{\frac{l}{L}} \left[ \frac{\tauN^{\mathrm{max}}}{\kfc \sigmaN} \left( \frac{4}{\pi} \frac{l}{L} - 1 \right) + \frac{2}{\pi \kfc} \frac{\FS}{\FN} \frac{L}{l} - 1 \right]
\label{eq:critical_length_simplist}
\end{equation}
where $l_{\mathrm{c}} = \sifIIc^2 / (\pi \tau_{\mathrm{d}}^2)$ is a characteristic interface length, and the dynamic interface traction is given by the friction law as $\tau_{\mathrm{d}} = \kfc \sigmaN = \kfc \FN / L$.

\begin{figure}
  \centering
  \includegraphics[width=0.99\columnwidth]{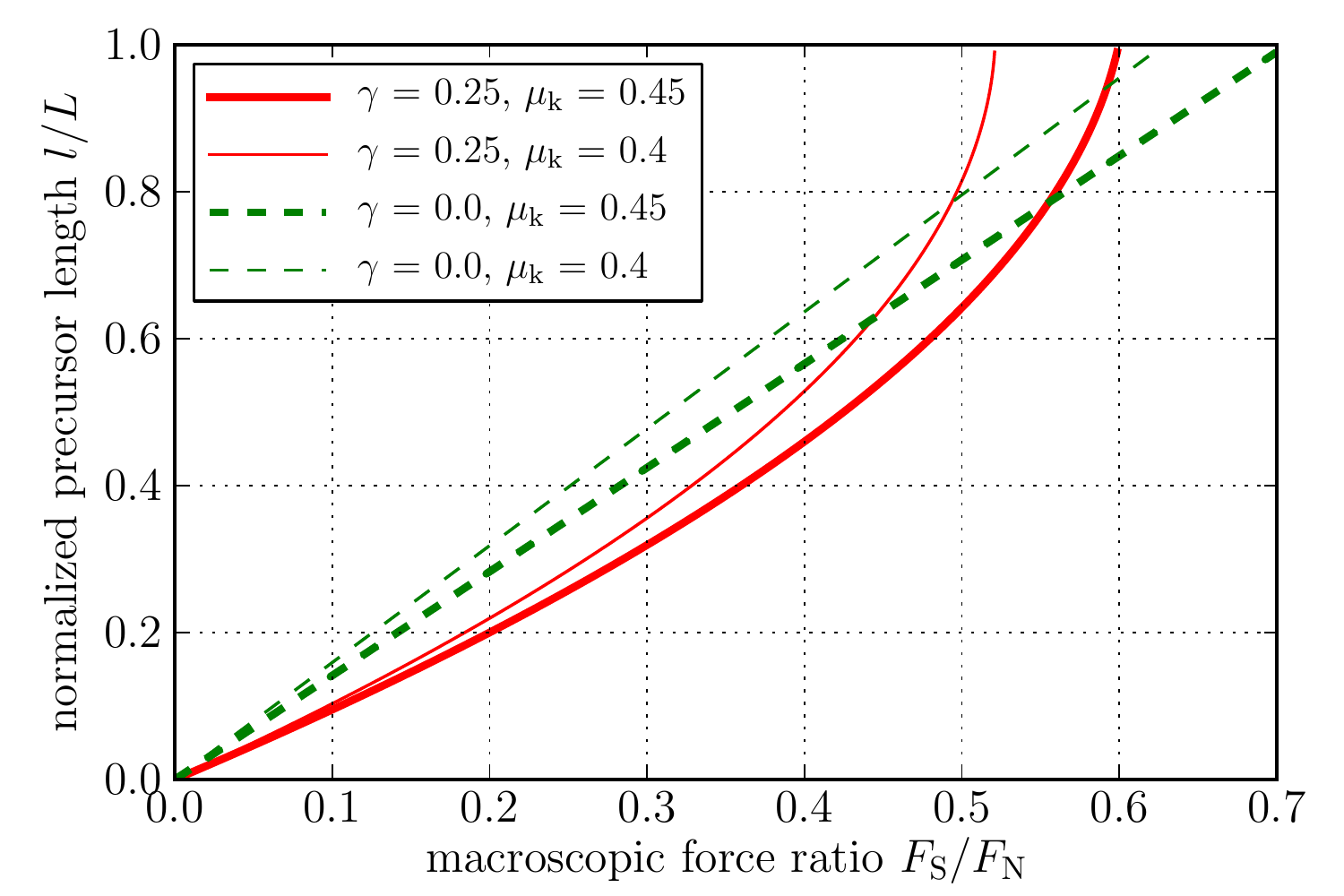}
  \caption{Precursor load-length relation of minimalistic model for various values of $\gamma = \tauN^{\mathrm{max}} / \sigmaN$ and $\kfc$. If the effect of frustrated Poisson's expansion is removed ($\gamma = 0$), then the precursor length increases linearly with the macroscopic force ratio. Generally, a smaller kinetic friction coefficient leads to longer precursors for a given macroscopic shear load.}
  \label{fig:minimalistic}
\end{figure}

The characteristic length is in our parameter domain much smaller than the interface length. Therefore, we can write $\sqrt{l_c/L} \approx 0$, and Eq.~\ref{eq:critical_length_simplist} becomes
\begin{equation}
  \frac{\FS}{\FN} = \frac{\pi \kfc}{2} \frac{l}{L} + \frac{\tauN^{\mathrm{max}}}{\sigmaN}\frac{l}{L} \left(\frac{\pi}{2} - 2 \frac{l}{L}\right) ~.
  \label{eq:scaling_simplist}
\end{equation}
With this simplifications, the macroscopic force ratio depends non-linearly on the normalized precursor length, and is controlled by two parameters: the kinetic friction coefficient $\kfc$ and the Poisson's expansion effect $\tauN^{max}/\sigmaN$. Note that this prediction neglects all stress redistributions, as in models A and AB.  The evolution of $l/L \, - \, \FS/\FN$  from Eq.~\ref{eq:scaling_simplist} is shown in Fig.~\ref{fig:minimalistic} with $\tauN^{\mathrm{max}}/\sigmaN=0.25$ and $\mu_k=0.45$ [thick red curve]. The scaling compares well with its equivalent of model A [see Fig.~\ref{fig:results}(c)]. If shear tractions due to frustrated Poisson's expansion are eliminated ($\tauN^{\mathrm{max}} = 0$),  Eq.~\ref{eq:scaling_simplist} predicts a linear load-length relation with a proportionality factor $\pi\kfc/2$. This prediction [green curve in Fig.~\ref{fig:minimalistic}], can be related to the almost linear evolution of model AB. The kinetic friction coefficient $\kfc$ appears in this scaling, whereas other interface parameters were neglected through the assumption that the characteristic length is much smaller than the interface ($\l_{\mathrm{c}} \ll L$). As already observed for the full LEFM theory, its influence on the load-length curve is obvious [compare thin with thick lines in Fig.~\ref{fig:minimalistic}], and small values of $\kfc$ lead to longer precursors for given $\FS/\FN$.

\subsection{LEFM Prediction for Symmetric Set-up}
\label{sec:symmetric_set-up}

Up to this point, we have compared our model to existing experimental data, analyzed the influence of different material and interface parameters, and have studied the non-linearity of the precursor load-length relation. Now, we can use our LEFM model to predict the response of a different system for which no experimental data has been published yet.

\begin{figure}
  \includegraphics[width=0.99\columnwidth]{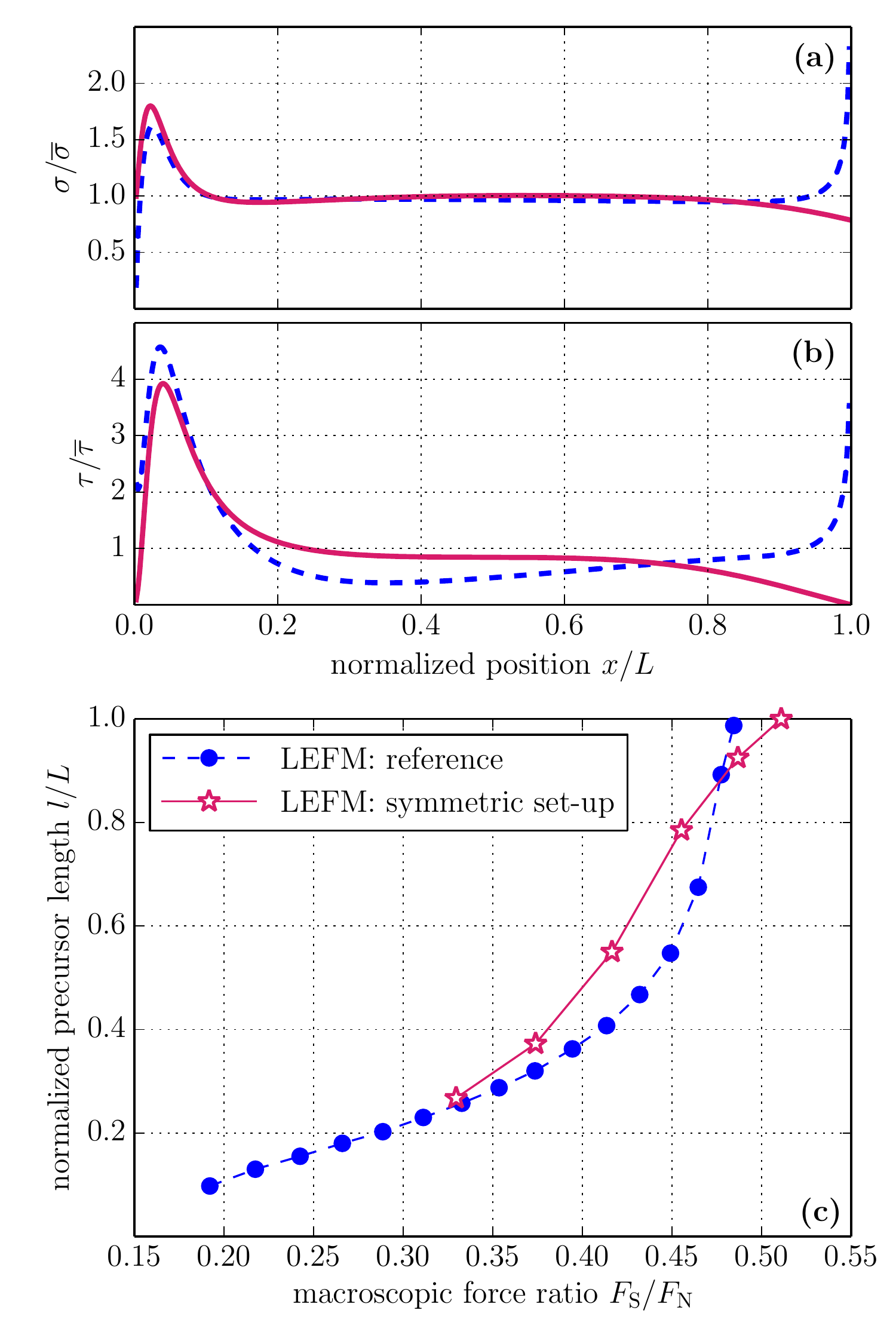}
  \caption{Comparison with a symmetric set-up, where the base has the same geometry as the slider. (a) Normalized contact pressure and (b) normalized shear traction of an unruptured interface for the reference set-up (blue dashed line), as also shown in Fig.~\ref{fig:initial_stresses}(d), and for a symmetric set-up (pink solid line) for $F_{N} = 5 F_{S}$. (c) Normalized precursor length $l/L$ reported with respect to the macroscopic force ratio $\FS/\FN$ for the reference set-up (blue dots), as also shown in Fig.~\ref{fig:results}, and for the symmetric set-up (pink stars).}
  \label{fig:symmetric}
\end{figure}

The set-up studied so far consists of a thin slider on a thicker base, which presents characteristics of a bi-material interface due to differences in the effective stiffness. This bi-material property influences the rupture propagation \cite{weertman:1980}.  It is potentially interesting to remove this effect from experimental observations of frictional precursors by using a set-up with a single-material interface.
We thus consider a symmetric system, where the base has the same geometry as the slider (in all three directions) and provide first insights to the propagation of precursors along a single-material interface. In this system, the non-zero $\tau_N$ due to frustrated Poisson's expansion [see Fig.~\ref{fig:initial_stresses}(b)] is naturally eliminated. Also all other interface traction components are different in a symmetric set-up and are computed with additional static finite-element simulations. An example of an effective normalized contact pressure of the symmetric set-up is shown in Fig.~\ref{fig:symmetric}(a) [solid pink line] and compared with the normalized contact pressure of the reference set-up [dashed blue line], which was already reported in Fig.~\ref{fig:initial_stresses}(d). As expected, the main difference is the absence of the edge singularity in the symmetric set-up at $x/L > 0.9$, which will only have a small influence on the precursor mechanics. In Fig.~\ref{fig:symmetric}(b), the normalized shear traction at the interface of the symmetric set-up [solid pink line] and the reference set-up [dashed blue line] are compared. The symmetric set-up is generally a system of lower stiffness, which leads to a peak at approximately $x/L = 0.05$ which is smaller than in the reference set-up but a stress level that is considerably higher up to $x/L = 0.7$.

The precursor load-length prediction of our LEFM model for the symmetric set-up is shown in Fig.~\ref{fig:symmetric}(c) [pink stars] and compared with the prediction for the reference system [blue dots], which was already shown in Fig.~\ref{fig:results}. In the symmetric set-up, the first precursor appears at higher macroscopic force ratio, which is the result of the lower stress peak in $\tau_r(x)$ [see Fig.~\ref{fig:symmetric}(b)]. The length of the first precursor is about the same than the precursor propagating at the same $\FS/\FN$ in the reference system. However, the precursor lengths increase faster in the symmetric system and the load-length relation presents an inflection point between the third and fourth precursors. Moreover, there are considerably less precursors in the prediction for the symmetric set-up [precursor length increments are larger], which indicates that it is harder to experimentally observe precursors in such a system.

\section{Conclusion}
\label{sec:conclusion}

We showed that a theoretical model based on linear elastic fracture mechanics predicts quantitatively well the precursor behavior observed in laboratory experiments~\cite{rubinstein:2007}. Using this model, we showed that the kinetic friction coefficient is key to an accurate prediction of the precursor length as it directly affects the stress intensity factor through the stress drop along the interface crack. Moreover, we showed that the variation of material parameters (within their uncertainty range) does not affect the observed precursor load-length relation. 
By simplifying this model in various ways, we analyzed different aspects that influence the non-linearity of the precursor growth and demonstrated that the shear tractions due to frustrated Poisson's expansion and the discrete nature of precursors are the main sources of the observed non-linearity.
The redistribution of the shear tractions along the interface caused by each precursor is essential to the load-length relation. With the results of this theoretical description of slip precursors, we provide evidence for considering frictional slip and precursors as a fracture phenomenon.

\begin{acknowledgements}
  The research described in this article is supported by the European Research Council (ERCstg UFO-240332), and the Swiss National Science Foundation (grant PMPDP2-145448). JPA was funded by US NSF (grant EAR-1015704).
\end{acknowledgements}

\bibliographystyle{spphys}       

\end{document}